\documentclass[tradiabstract]{aa}  

\usepackage{graphicx}
\usepackage{txfonts}
\usepackage{hyperref}
\usepackage{array}
\newcolumntype{P}[1]{>{\centering\arraybackslash}p{#1}}
\newcolumntype{M}[1]{>{\centering\arraybackslash}m{#1}}
\newcolumntype{N}{@{}m{0pt}@{}}

\usepackage{color}

\graphicspath{{figures/}}
\hypersetup{ bookmarks=true, unicode=true, pdftoolbar=true, pdfmenubar=true, pdffitwindow=true, pdfstartview={FitH}, pdfauthor={SJ}, colorlinks=true, linkcolor=blue,citecolor=blue, urlcolor=black, breaklinks }
 
\begin{document}

   \title{Star formation history from the cosmic infrared background anisotropies}
   
   \author{A. Maniyar
          \inst{1}
          \and
          M. B\'ethermin
          \inst{1}
          \and
          G. Lagache
         \inst{1}
          }

   \institute{ $^1$Aix Marseille Univ., CNRS, LAM, Laboratoire d'Astrophysique de Marseille, Marseille, France\\
        \email{\href{mailto:abhishek.maniyar@lam.fr}{\textrm{abhishek.maniyar@lam.fr}
        }}
        \label{inst1}                   
        }

   \date{Received December 19, 2017/ Accepted 29 January 2018}

   \abstract
{
We present a linear clustering model of cosmic infrared background (CIB) anisotropies at large scales that is used to measure the cosmic star formation rate density up to redshift 6, the effective bias of the CIB, and the mass of dark matter halos hosting dusty star-forming galaxies. This is achieved using the {\it Planck} CIB auto- and cross-power spectra (between different frequencies) and CIBxCosmic microwave background (CMB) lensing cross-spectra measurements, as well as external constraints (e.g. on the CIB mean brightness). We recovered an obscured star formation history which agrees well with the values derived from infrared deep surveys and we confirm that the obscured star formation dominates the unobscured formation up to at least z=4. The obscured and unobscured star formation rate densities are compatible at $1\sigma$ at z=5. We also determined the evolution of the effective bias of the galaxies emitting the CIB and found a rapid increase from $\sim$0.8 at z$=$0 to $\sim$8 at z$=$4. At 2$<$z$<$4, this effective bias is similar to that of galaxies at the knee of the mass functions and submillimetre galaxies. This effective bias is the weighted average of the true bias with the corresponding emissivity of the galaxies. The halo mass corresponding to this bias is thus not exactly the mass contributing the most to the star formation density. Correcting for this, we obtained a value of log(M$_h$/M$_{\odot}$)=12.77$_{-0.125}^{+0.128}$ for the mass of the typical dark matter halo contributing to the CIB at z=2. Finally,  using a Fisher matrix analysis we also computed how the uncertainties on the cosmological parameters affect the recovered CIB model parameters, and find that the effect is negligible.}


 \authorrunning{Maniyar, B\'ethermin, Lagache}
 \titlerunning{History of star formation from CIB anisotropies}
  \maketitle   
   

\section{Introduction}
\label{x1}
How and when the galaxies assembled their stars are two of the most pressing questions of modern cosmology. 
In the last decade, it has become clear that the dusty star-forming galaxies are significant contributors to this process as they pin down the episodes of galaxy-scale star formation   \citep[e.g.][]{Gispert_2000, Casey_2014}. Indeed they are a critical player in the assembly of stellar mass and the evolution of massive galaxies \citep[e.g.][]{Bethermin_2015}.\\
Dusty star-forming galaxies are difficult to detect individually at high redshift because they are so faint and numerous compared to the angular resolution achievable in the far-infrared to millimetre wavelengths that the confusion limits the ultimate sensitivity \citep[e.g.][]{Nguyen_2010}. As a result, single-dish experiments, such as \textit{Herschel} and \textit{Planck}, can only see the brightest objects that represent the tip of the iceberg in terms of star formation rates and halo masses. These experiments are nevertheless sensitive enough to measure  the cosmic infrared background (CIB),  the cumulative infrared emission from all galaxies throughout cosmic history \citep{ Lagache_2005, Planck_cib_2011}. The mean value of the CIB gives the amount of energy released by the star formation that has been reprocessed by dust \citep{Dole_2006}, while CIB anisotropies trace the large-scale distribution of dusty star-forming galaxies, and can thus be used to trace the underlying distribution of the dark matter halos in which such galaxies reside \citep{Amblard_2011, Viero_2013, Bethermin_2013, Planck_cib_2014}. \\

The cosmic history of star formation is one of the most fundamental observables in astrophysical cosmology. 
A consistent picture has emerged in recent years, whereby the star-formation rate density peaks at z$\sim$2, and declines exponentially at lower redshift. The Universe was much more active in forming stars in the past; the star formation rate density was about ten times higher at $z\sim$2 than is seen today \citep{Madau_2014}. These highest star formation rate densities are the consequence of the rising contribution from the dusty star formation. At higher redshift ($z>>2$), the level of dust-obscured star formation density is still a matter of debate; there is no census for the star formation rate density (SFRD) selected from dust emission alone \citep[e.g.][]{Koprowsky_2017}.  With their unmatched redshift depth, CIB anisotropies can be used to improve our understanding of early star formation. The contribution of dusty galaxies to the SFRD can be derived from their mean emissivity per comoving unit volume as derived from CIB anisotropy modelling. This has been investigated by \cite{Planck_cib_2014}. In this paper, we revisit this determination using {\it Planck} CIB measurements combined with the latest observational constraints and theoretical developments.\\

Galaxy clustering is another piece of  information embedded into the CIB anisotropies \citep{Knox_2001, Lagache_2007}. The distribution of galaxies is known to be biased relative to that of the dark matter. In the simplest model, on large scales, it is assumed that the galaxy and dark matter mass-density fluctuations are related through a constant bias factor $b$ such that $(\Delta \rho / \rho)_{light}$ = $ b (\Delta \rho / \rho)_{mass}$. Light traces mass exactly if $b=1$ (which is the case in the local Universe, on average and at scales larger than the individual galaxy halos); if $b>1$, light is a biased tracer of mass, as expected if the galaxies form in  the highest peaks of the density field. 
 The galaxy bias is dependent on the host dark matter halo mass and the redshift \citep{Mo_1996}. As expected by theory \citep[e.g.][]{Kaiser_1986, Wechsler_1998} and confirmed by observations \citep[e.g.][]{Adelberger_2005}, the biasing becomes more pronounced at high redshift. As most of the star formation over the history of the Universe has been obscured by dust  \citep{Dole_2006},  studying the clustering of dusty galaxies  is crucial to exploring the link between dark matter halo mass and star formation.  Measurements of correlation function of  individually  resolved  galaxies, whilst not forming  the dominant contribution to the CIB, give constraints on the host halo mass of the brightest star-forming objects, the submillimetre galaxies (SMGs). \citet{Maddox_2010} and \citet{Cooray_2010} measured the correlation function of \textit{Herschel}/SPIRE galaxies (see also \citealt{weiss_2009} and \citealt{Williams_2011} for measurements from ground-based observations), but their results were inconsistent probably because of the difficulty in  building well-controlled selections in low angular resolution data \citep{Cowley_2016}. One way to get more accurate measurements is to use the cross-correlation with optical/near-IR samples \citep{Hickox_2012,Bethermin_2014,Wilkinson_2017}. These analyses found the host halo mass of galaxies with the star formation rate (SFR)$>$100\,M$_\odot$ to be $\sim$10$^{12.5-13}$\,M$_\odot$. Reaching the population that contributes to the bulk of the CIB requires  modelling the angular power spectrum of CIB. Making various assumptions such as the form of the relationship between galaxy luminosity and halo mass, a typical halo mass for galaxies that dominate the CIB power spectrum has been found to lie between $10^{12.1\pm0.5}$ M$_{\odot}$ and $10^{12.6\pm0.1}$ M$_{\odot}$ \citep{Viero_2013, Planck_cib_2014}. In this paper, we exploit the strength of {\it Planck}, which is a unique probe of the large-scale anisotropies of the CIB. We use only the linear part of the power spectra, and thus consider a simple model (with few parameters), to measure the galaxy clustering of dusty star-forming galaxies up to $z\sim$5.\\

All previous CIB angular power spectrum analyses have considered a fixed cosmology. Assuming a cosmology is a fair assumption as the model of the CIB galaxy properties has far greater uncertainties than on the cosmological parameters \citep{Planck_cosmo_2016}. However, the CIB power spectrum depends on key cosmological ingredients, such as the dark matter power spectrum, cosmological distances, and cosmological energy densities. Therefore, it is worth studying the impact of the cosmology on CIB modelling parameters, which we  do here for the first time.\\

The paper is organised as follows. We describe the CIB anisotropy modelling and the data we  use to constrain the model in Sect.\,\ref{sec:cib_model}. In Sect.\,\ref{sec:CIB_z_SFRD}, we present and discuss the redshift distribution of CIB anisotropies and mean level, and the star formation rate density from z=0 to z=6 that are obtained by fitting our model to the CIB and CIBxCMB lensing power spectra measurements and by using extra constraints coming from galaxy number counts and luminosity functions. We then discuss the host dark matter halo mass of CIB galaxies in Sect.\,\ref{sec:DMH_Mass} by first measuring the effective bias $b_{eff}$, then converting it to the mean bias of halos hosting the dusty galaxies, before finally computing the host dark matter halo mass of the dusty star formation. In Sect.\,\ref{sec:cosmo_cib}, we study the effect of cosmology on the CIB. We conclude in Sect.\,\ref{sec:cl}.\\
Throughout the paper, we used a Chabrier mass function \citep{Chabrier_2003} and the Planck 2015 flat $\Lambda$CDM cosmology \citep{Planck_cosmo_2016} with $\Omega_m = 0.33$ and $H_0=67.47$\,km\,s$^{-1}$\,Mpc$^{-1}$.

\section{CIB modelling on large scales}
\label{sec:cib_model}
\subsection{CIB power spectrum}
The measured CIB intensity $I_\nu$ at a given frequency $\nu$, in a flat Universe,  is given by
\begin{equation}
I_\nu =  \int \dfrac{d\chi}{dz} a j(\nu,z) dz \, ,
\end{equation}
where $j$ is the comoving emissivity, $\chi(z)$ is the comoving distance to redshift $z$, and $a = 1/(1+z)$ is the scale factor. 
The angular power spectrum of CIB anisotropies is defined as
\begin{equation}
C_l^{\nu \times \nu'} \times \delta _{ll'}\delta_{mm'}= \Big \langle \delta I_{lm}^{\nu} \delta I_{l'm'}^{\nu'} \Big \rangle \,.
\end{equation}
Combining these two equations and using the Limber approximation, we get
\begin{equation}
C_l^{\nu \times \nu'} = \int \dfrac{dz}{\chi^2} \dfrac{d\chi}{dz}a^2 \bar{j}(\nu,z) \bar{j}(\nu',z)P_j^{\nu \times \nu'}(k = l/\chi,z) \, ,
\end{equation}
where $P_j^{\nu \times \nu'}$ is the 3D power spectrum of the emissivities and is defined as
\begin{equation}
\Big \langle \delta j(k,\nu) \delta j(k',\nu') \Big \rangle = (2\pi)^3 \bar{j}(\nu)\bar{j}(\nu')P_j^{\nu \times \nu'}(k)\delta^3(k-k')\, ,
\end{equation}
with $\delta j$ being the emissivity fluctuation of the CIB. In our analysis, we are interested in modelling the CIB anisotropies on large angular scales ($\ell < 800$), where the clustering  dominated by the correlation between dark matter halos and the non-linear effects can be neglected. We can then model the CIB anisotropies equating $P_j$ with $P_{gg}$, which is the galaxy power spectrum. On large angular scales, $P_{gg}(k,z) = b_{eff}^2(z)P_{lin}(k,z)$. Here, $b_{eff}$ is the effective bias factor for dusty galaxies at a given redshift, i.e. the mean bias of dark matter halos hosting dusty galaxies at a given redshift weighted by their contribution to the emissivities \citep[see][]{Planck_cib_2014}. The fact that more massive halos are more clustered and have host galaxies emitting more far-infrared emission is implicitly taken into account through this term. We consider $b_{eff}$ as scale independent, which is a good approximation at the scales we are interested in. Here,  $P_{lin}(k,z)$ is the linear theory dark matter power spectrum. Finally, the linear CIB power spectrum is given as
\begin{equation}\label{eq:clcib}
C_l^{\nu \times \nu'} = \int \dfrac{dz}{\chi^2} \dfrac{d\chi}{dz}a^2 b_{eff}^2 \bar{j}(\nu,z) \bar{j}(\nu',z)P_{lin}(k = l/\chi,z)\,.
\end{equation}
This is known as the linear model of CIB anisotropies.
The emissivities $\bar{j}(\nu,z)$ (in Jy/Mpc) are derived using the star formation rate density $\rho_{SFR}$ (M$_{\odot}$yr$^{-1}$Mpc$^{-3}$) following
\begin{equation}\label{eq:jnu}
\bar{j}(\nu,z) = \dfrac{\rho_{\textrm{SFR}}(z)(1+z)S_{\nu,eff}(z)\chi^2(z)}{K}
,\end{equation}
where $K$ is the Kennicutt constant (\citealp{Kennicutt_1998}) and $S_{\nu,eff}(z)$ is the mean effective spectral energy distribution (SED) for all the dusty galaxies at a given redshift. We used a Kennicutt constant corresponding to a Chabrier initial mass function ($ \textrm{SFR}/L_{IR} = 1.0 \times 10^{-10} \textrm{M}_{\odot}\textrm{yr}^{-1}$). Here,  $S_{\nu,eff}(z)$ (in Jy/L$_{\odot}$) is the average SED of all galaxies at a given redshift weighted by their contribution to the CIB. They have been computed using the method presented in \citet{Bethermin_2013}, but assuming the new updated SEDs calibrated with the \textit{Herschel} data and presented in \citet{Bethermin_2015} and \citet{Bethermin_2017}. Compared to the previous SEDs used in the \citet{Bethermin_2013} and the \citet{Planck_cib_2014}, the dust in these new SED templates is warmer at z$>$2. We used CAMB\footnote{\url{http://camb.info/}} to generate cold dark matter power spectra $P_{lin}(k)$ for given redshifts. For the effective bias, we chose the following parametric form based on redshift evolution of the dark matter halo bias: 
\begin{equation} \label{eq:beff}
b_{eff} (z)= b_0 + b_1z + b_2z^2 \,.
\end{equation}
We also tested alternative parametric forms for the effective bias where the evolution of the bias is slower compared to the above equation. The results are presented in the Appendix. \\
To describe $\rho_{SFR}$, we used the parametric form of the cosmic star formation rate density proposed by \cite{Madau_2014} and given by
\begin{equation}
\rho_{\textrm{SFR}}(z) = \alpha\dfrac{(1+z)^{\beta}}{1+ {[ (1+z)/{\gamma} ]}^{\delta}} \textrm{M}_{\odot} \textrm{year}^{-1} \textrm{Mpc}^{-3}\, ,
\end{equation}
where $\alpha$, $\beta$, $\gamma$, and $\delta$ are free parameters in our CIB model. 

\begin{table*}[t]
\centering
\begin{tabular}{cccc}
\hline
\hline
Frequency &  Conversion factor & Abs. calibration errors & PR1/PR2\\\relax
[GHz] & $\textrm{MJy}sr^{-1}$ $[\nu I_\nu = \textrm{const}]/K_{CMB}$ & [\%] & \\ 
\hline
 100      & $244.1\pm{0.3}$ & 0.09$^a$ & 0.994$^c$ \\
 143       & $371.74\pm{0.07}$ & 0.07$^a$ & 0.993$^c$\\
 217      & $483.69\pm{0.012}$ & 0.16$^a$ & 0.991$^c$\\
 353      & $287.45\pm{0.009}$ & 0.78$^a$ & 0.977$^c$\\
 545        & $58.04\pm{0.03}$ & 3.1$^b$ & 1.018$^a$\\
 857        & $2.27\pm{0.03}$ & 3.4$^b$ & 1.033$^a$\\
 3000        & -- & 13.5 & -- \\
 \hline
 \end{tabular}\\
\tablefoot{{\tiny\tablefoottext{a}{from \cite{planck_HFI_2016}}. \tablefoottext{b}{from \cite{Planck_JLP_2016}}. \tablefoottext{c}{computed using exactly the same method as that used in  \cite{planck_HFI_2016}}.}}
\caption{Useful numbers used in our analysis. The first column gives the unit conversion factors between MJy\,sr$^{-1}[\nu I_\nu = \textrm{const}]$ and K$_{CMB}$; the second column gives the error on the absolute calibration; and the third column gives the factors we applied to the {\it Planck} CIB measurements to account for a more accurate absolute calibration between the two releases of the {\it Planck} data (PR1 and PR2).}     \label{tab:1}
\end{table*}

\subsection{CIB-CMB lensing} 
Large-scale distribution of the matter in the Universe gravitationally deflects the CMB photons which are propagating freely from the last scattering surface. This gravitational lensing leaves imprints on the temperature and polarisation anisotropies. These imprints can be used to reconstruct a map of the lensing potential along the line of sight \citep{Okamoto_2003}. Dark matter halos located between us and the last scattering surface are the primary sources for this CMB lensing potential \citep{Lewis_2006} and it has been shown \citep[e.g.][]{Song_2003} that a strong correlation between CIB anisotropies and a lensing derived projected mass map is expected. \\
We calculated the cross-correlation between the CIB and CMB lensing potential \citep{Planck_ciblensing_2014}, which is given by
\begin{equation} \label{eq:lens}
C_l^{\nu\phi} = \int b_{eff} \bar{j}(\nu,z) \dfrac{3}{l^2}\Omega_m H_0^2 \bigg(\dfrac{\chi_* - \chi}{\chi_*\chi}\bigg) P_{lin}(k = l/\chi,z) d\chi \,,
\end{equation}
where $\chi_*$ is the comoving distance to the CMB last scattering surface, $\Omega_m$ is the matter density parameters, and $H_0$ is the value of the Hubble parameter today. From Eq.\,\ref{eq:lens}, we see that $C_l^{\nu,\phi}$ is proportional to $b_{eff}$, whereas $C_l^{CIB}$ is proportional to $b_{eff}^2$. Therefore, also using  the CIB-CMB lensing potential measurement when fitting the CIB model helps us resolve the degeneracy between the evolution of $b_{eff}$ and $\rho_{\textrm{SFR}}$ to some extent. 

\subsection{Constraints on the model through data}
\subsubsection{Observational constraints on the power spectra} \label{subsub:obs}
We used the CIB angular power spectra  measured by \cite{Planck_cib_2014}. The measurements were obtained by cleaning the {\it Planck} frequency maps from the CMB and galactic dust, and they were further corrected for SZ and spurious CIB contamination induced by the CMB template, as discussed in \cite{Planck_cib_2014}. We used the measurements at the four highest frequencies (217, 353, 545, and 857\,GHz from the HFI instrument). For the 3000 GHz, far-infrared data from IRAS (IRIS, \citealp{Miville-Deschenes_2005}) were used. 
As we are using the linear model we fit the data points only for $l \leqslant 600,$ which are dominated by the 2-halo term (>90\%, \citealp{Bethermin_2013, Planck_cib_2014}). 

CIB - CMB lensing potential cross-correlation values and error bars are available for the six Planck HFI channels (100, 143, 217, 353, 545, and 857\,GHz) and are provided in \cite{Planck_ciblensing_2014}. These values range from $\ell = 163$ to $\ell = 1937$. As discussed in \cite{Planck_ciblensing_2014}, the non-linear term can be neglected in this range of multipoles.

We used the $\nu I_\nu = \mathrm{const}$ photometric convention. Thus, the power spectra computed by the model need to be colour-corrected from  our CIB SEDs to this convention. Colour corrections are 1.076, 1.017, 1.119, 1.097, 1.068, 0.995, and 0.960 at 100, 143, 217, 353, 545, 857, and 3000\,GHz, respectively, \citep{Planck_cib_2014}. The CIB power spectra are then corrected as  
\begin{equation}
\label{eq:cc}
C_{l,\nu,\nu '}^{\textrm{model}}\: \textrm{x}\: \textrm{cc}_\nu \: \textrm{x}\: \textrm{cc}_{\nu'} = C_{l,\nu,\nu '}^{\textrm{measured}}\,.
\end{equation}

Calibration uncertainties are not accounted for in the CIB power spectra error bars and are treated differently. \cite{Bethermin_2011} introduce a calibration factor $f_\textrm{cal}^\nu$ for galaxy number counts. We used  a similar approach here. We put Gaussian priors on these calibration factors for different frequency channels with an initial value of 1 and the error bars as given in Table\,\ref{tab:1}. The CIB measurements from {\it Planck} were obtained with the PR1 release of the maps. Between the PR1 and PR2 releases, the absolute calibration  improved further, and we thus wanted to consider the absolute calibration of the PR2 release for the CIB measurements. Consequently, the CIBxCIB and CIBxCMB\,lensing power spectra data points were corrected for the absolute calibration difference of the two releases (following the numbers given in  Table\,\ref{tab:1}), and we used the absolute calibration errors as given for the PR2 release \citep{planck_HFI_2016}.

\begin{figure*}[ht!]
\centering
\includegraphics[width=\textwidth]{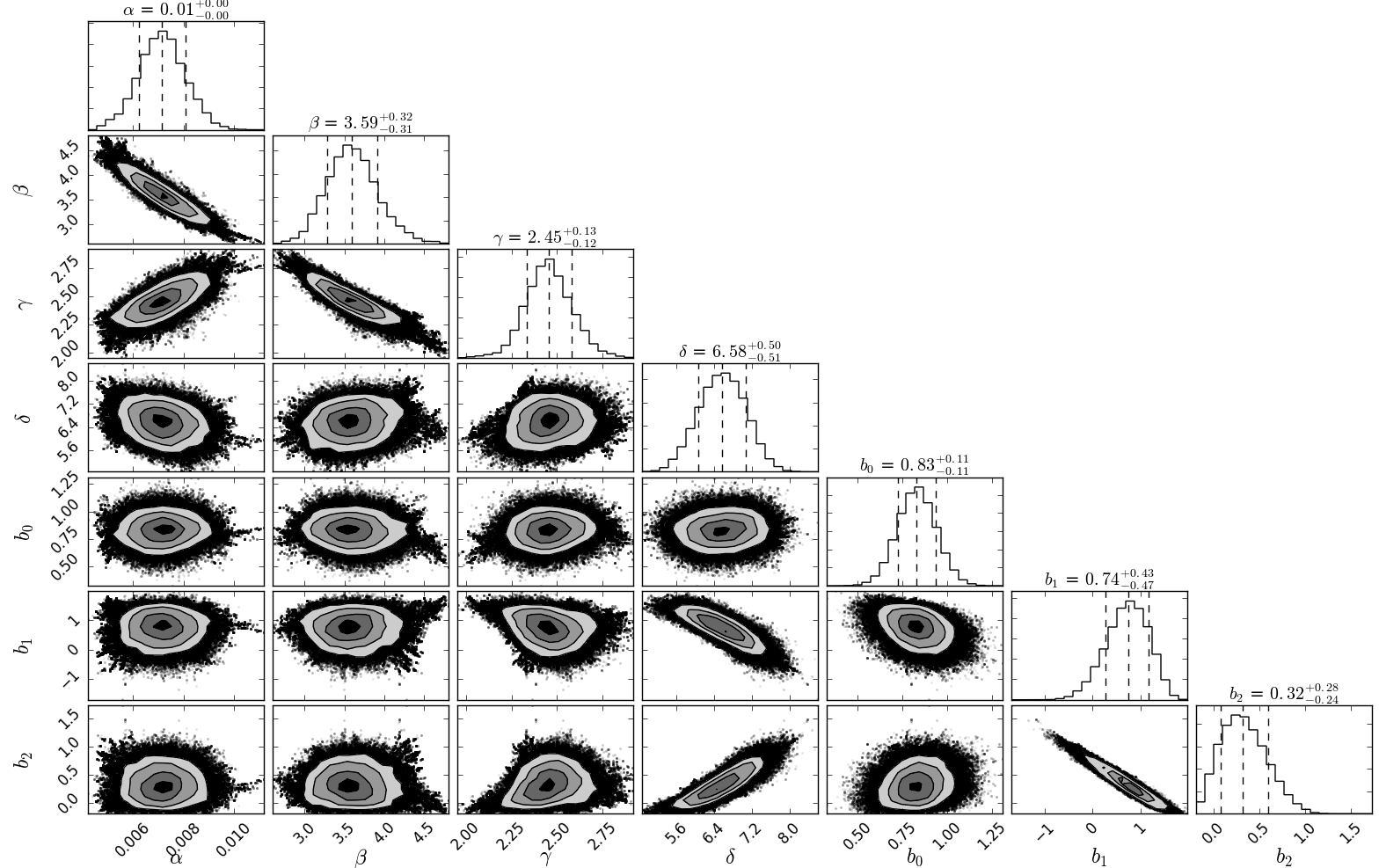}
\centering \caption{Posterior confidence ellipses for the CIB model parameters. All the calibration parameters lie within the $1\sigma$ range of their prior values. There is no significant degeneracy between any of the calibration parameters and CIB parameters and hence they are not shown here.} 
\label{fig:contours}
\end{figure*}

\subsubsection{External observational constraints} \label{subsub:extobs}

To put better constraints on $\rho_{\textrm{SFR}}$ and $b_{eff}$ parameters, in addition to the CIBxCIB and CIBxCMB\,lensing angular power spectra, we used some external observational constraints on the star formation rate density at different redshifts, local bias of the dusty galaxies, and mean CIB levels at different frequencies, as detailed below:
\begin{enumerate}
\item
We used the $\rho_{\textrm{SFR}}$ measurements at different redshifts that were obtained by measuring the IR luminosity functions from \cite{Gruppioni_2013, Magnelli_2013, Marchetti_2015} (see the discussion in Sect.\,\ref{sec:SFRD}).
This helped us to put better constraints on the parameters. The cosmological parameters used in these studies are different from the ones  used here. As mentioned before, we also perform a Fisher matrix analysis to study the effect of the CIB parameters on cosmology. For this purpose, we needed to convert all the observational $\rho_{\textrm{SFR}}$ data points to actual measurements which are cosmology independent. We thus used the observed flux in the range 8--1000\,$\mu$m (rest frame) per redshift bin per solid angle $\dfrac{d\textrm{B}_{\textrm{IR}}}{dzd\Omega}$. We perform the conversion as follows: 
\begin{equation}
\begin{split}
\dfrac{d\textrm{B}_\textrm{IR}}{dzd\Omega} & = \dfrac{d(\textrm{P}_\textrm{IR}/4\pi \textrm{D}_L^2)}{dzd\Omega} \\
    & = \dfrac{1}{4\pi \textrm{D}_L^2} \dfrac{d\textrm{P}_\textrm{IR}}{d\textrm{V}_c} \dfrac{d\textrm{V}_c}{dzd\Omega} \\
    & = \dfrac{1}{4\pi \textrm{D}_L^2} \dfrac{\rho_\textrm{SFR}}{K} \dfrac{D_H (1+z)^2 D_A^2}{E(z)} \,.
\end{split}
\end{equation}
Here $\dfrac{d\textrm{P}_\textrm{IR}}{d\textrm{V}_c}$ is the power emitted in the infrared per unit co-moving volume of space, $D_L$ is the luminosity distance, $D_A$ is the angular diameter distance, $D_H$ is the Hubble distance given by $c/H_0$ (with c and $H_0$ being the speed of light and Hubble's constant, respectively), and $E(z) = \sqrt{\Omega_m(1+z)^3 +\Omega_k (1+z)^2 + \Omega_\Lambda}$. This equation can  be simplified further using the relation $D_L = (1+z)^2 D_A$ and becomes
\begin{equation} \label{eq:flux}
\dfrac{d\textrm{B}_\textrm{IR}}{dzd\Omega} = \dfrac{1}{4\pi} \dfrac{\rho_\textrm{SFR}}{K} \dfrac{D_H (1+z)^2}{(1+z)^2E(z)} \,.
\end{equation} 
We also needed to convert the measurements to the same IMF; e.g. \cite{Gruppioni_2013, Magnelli_2013, Marchetti_2015} used the Salpeter IMF, whereas we  used the Chabrier IMF. Finally, we used these observed flux values $d\textrm{B}_{\textrm{IR}}/dzd\Omega$ in our fitting rather than the $\rho_\textrm{SFR}$ values. This conversion to cosmology-independent variables is particularly important for the Fisher matrix analysis presented in Sect.\,\ref{sec:cosmo_cib}, else we would wrongly find that CIB anisotropies can be used to significantly improve the uncertainties on the cosmological parameters.

\item
\cite{Saunders_1992} provide the fixed value for the product of the local bias of infrared galaxies $b_I$ and the cosmological parameter $\sigma_8$: $b_I \sigma_8 = 0.69 \pm 0.09$. We put a constraint of $b = 0.83 \pm 0.11$ on local bias value which has been converted using the $\sigma_8$ measured by \cite{Planck_cosmo_2016}.
\item
The mean level of the CIB at different frequencies has been deduced from galaxy number counts. The values we used are given in Table\,\ref{tab:mean_cib}. 
Similarly to what is being done for $C_{l,\nu,\nu '}$ (Eq.\,\ref{eq:cc}), the CIB computed by the model needs to be colour-corrected, from our CIB SEDs to the photometric convention of PACS, SPIRE, and SCUBA2. The colour corrections are computed using the \cite{Bethermin_a_2012} CIB model and the different bandpasses and are given in Table\,\ref{tab:mean_cib}.
\item Finally, we also put physical constraints on $\rho_{SFR}$ and $b_{eff}$ parameters such that the star formation rate and galaxy bias is positive at all redshifts. 
\end{enumerate}

\begin{table}[h]
 \centering
\begin{tabular}{cccc}
\hline
\hline 
Instrument & \begin{tabular}{@{}c@{}}Freq \\ GHz\end{tabular} & \begin{tabular}{@{}c@{}}$\nu \textrm{I}_\nu$ \\ $\textrm{nW}\textrm{m}^{-2}\textrm{sr}^{-1}$\end{tabular}   &  \begin{tabular}{@{}c@{}}Colour\\ correction\end{tabular} \\ 
\hline
 \begin{tabular}{@{}c@{}}Herschel/PACS\\ \cite{Berta_2011}\end{tabular}   & 3000 & $12.61\substack{+8.31 \\ -1.74}$ & 0.9996 \\ [6pt]
 \begin{tabular}{@{}c@{}}Herschel/PACS\\ \cite{Berta_2011}\end{tabular}   & 1875 & $13.63\substack{+3.53 \\ -0.85}$ & 0.9713 \\ [6pt]
 \begin{tabular}{@{}c@{}}Herschel/SPIRE\\ \cite{Bethermin_2012}\end{tabular}   & 1200 & $10.1\substack{+2.60 \\ -2.30}$ & 0.9880 \\ [6pt]
 \begin{tabular}{@{}c@{}}Herschel/SPIRE\\ \cite{Bethermin_2012}\end{tabular}   & 857 & $6.6\substack{+1.70 \\ -1.60}$ & 0.9887 \\ [6pt]
 \begin{tabular}{@{}c@{}}JCMT/SCUBA2\\ \cite{Wang_2017}\end{tabular}   & 667 & $1.64\substack{+0.36 \\ -0.27}$ & 1.0057 \\ [6pt]
 \begin{tabular}{@{}c@{}}Herschel/SPIRE\\ \cite{Bethermin_2012}\end{tabular}   & 600 & $2.8\substack{+0.93 \\ -0.81}$ & 0.9739 \\ [6pt]
 \begin{tabular}{@{}c@{}}JCMT/SCUBA2\\ \cite{Zavala_2017}\end{tabular}   & 353 & $0.46\substack{+0.04 \\ -0.05}$ & 0.9595 \\ [6pt]
 \begin{tabular}{@{}c@{}}ALMA\\ \cite{Aravena_2016}\end{tabular}   & 250 & $>0.08$ & - \\ [6pt]
 \hline
\end{tabular}
\newline
\centering \caption{Mean levels of CIB from different instruments at different frequencies with the corresponding colour corrections used to convert them to the {\it Planck} and {\it IRAS} bandpasses.}
\label{tab:mean_cib}
\end{table}

\begin{figure*}
\centering
\includegraphics[width=\textwidth,height=15cm]{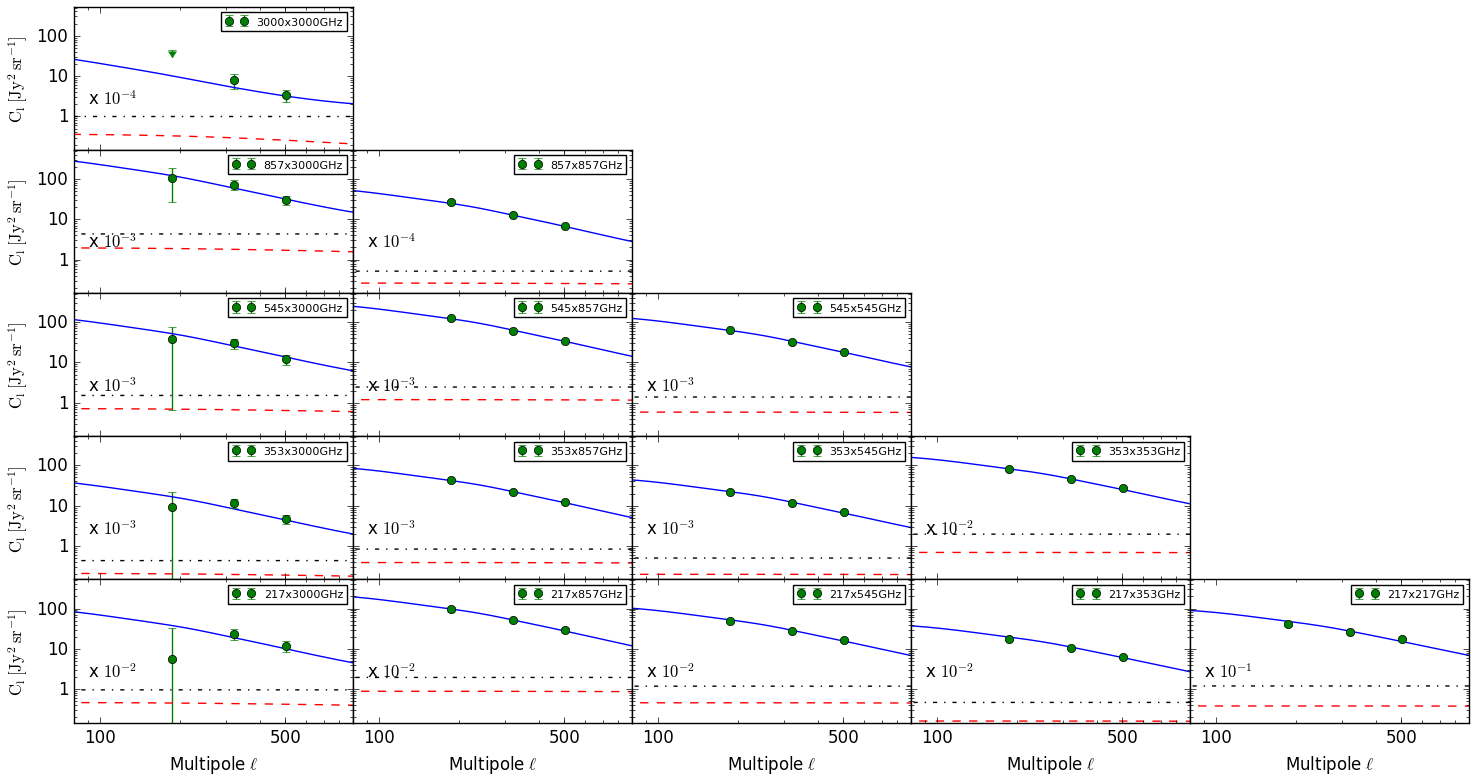}
\centering \caption{Measurements of the CIB auto- and cross-power spectra obtained by \textit{Planck} and \textit{IRAS} (extracted from \citep{Planck_cib_2014}) and the best fit CIB linear model. Data points on higher $\ell$ are not shown as they are not used for fitting. We also show the shot noise (black dash-dotted line) and one-halo term (red dashed line) for all the frequencies.}
\label{fig:cibfit}
\end{figure*}

\subsubsection{Fitting the data} \label{subsub:fitting}
We performed a  Markov chain Monte Carlo (MCMC) analysis on the global CIB parameter space. Python package `emcee' \citep{emcee} is used for this purpose. We have a 12-dimensional parameter space: $\{\alpha, \beta, \gamma, \delta, b_0, b_1, b_2, f_{217}^\textrm{cal}, f_{353}^\textrm{cal}, f_{545}^\textrm{cal}, f_{857}^\textrm{cal}, f_{3000}^\textrm{cal}\}$. Our global $\chi^2$ has a contribution from the CIBxCIB and CIBxCMB\, lensing angular power spectra measurements, priors on calibration factors, and priors imposed by the external observational constraints mentioned above. We assumed Gaussian uncorrelated error bars for measurement uncertainties. The 1-halo and Poisson terms have very little measurable contribution (less than 10\%) at $l \leq 600$, where we are fitting the data points. Similar to the procedure used in \cite{Planck_cib_2014}, we added the contribution of these terms derived from the \cite{Bethermin_2013} model to our linear model.

\subsection{Results}
\begin{figure}
\centering
\includegraphics[width=9cm,height = 10cm]{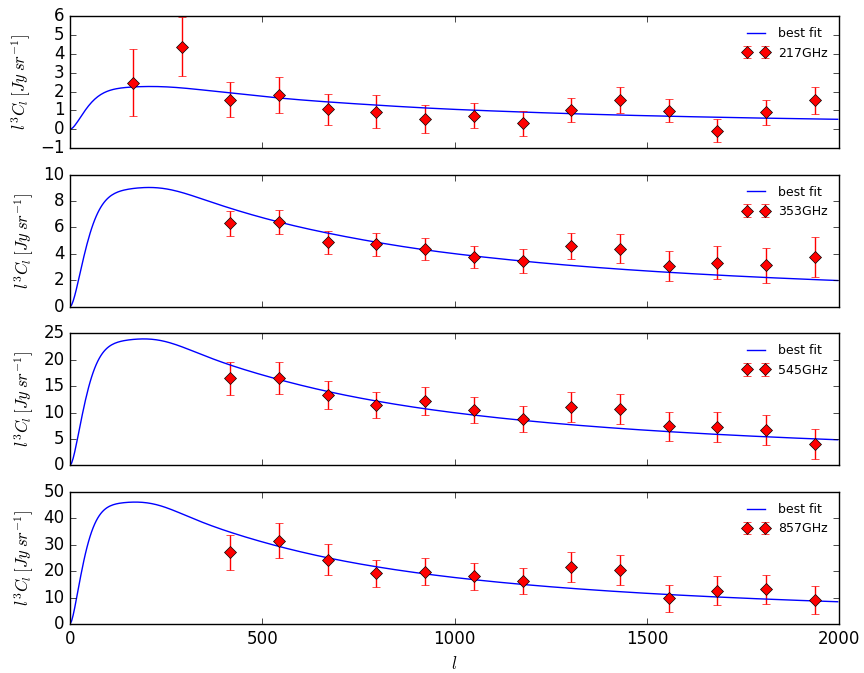}
\centering \caption{CIBxCMB\,lensing cross-power spectra. Our best fit model is shown by the blue curves for different frequency channels. Measurements (red data points) are from \cite{Planck_ciblensing_2014}. They have been included in the likelihood to calculate and get better constraints on the CIB linear model.}
\label{fig:lens_fit}
\end{figure}

We present the results from our fit in Table \ref{tab:3}. We find a good fit for the model and the posterior of all the parameters with a Gaussian prior (local effective bias and calibration factors) are within a $1\sigma$ range of the prior values. The $1\sigma$ and $2\sigma$ confidence regions for the $\rho_{SFR}$ and $b_{eff}$ parameters are shown in Fig.\,\ref{fig:contours}. As expected, we observe strong degeneracies between $\rho_\textrm{SFR}$ and $b_\textrm{eff}$, and also within the $\rho_\textrm{SFR}$ and $b_\textrm{eff}$ parameters themselves. Figures\,\ref{fig:cibfit} and \ref{fig:lens_fit} show the fit for the linear model to the observational data points for all CIBxCIB auto- and cross-power spectra, and CIBxCMB\,lensing power spectra, respectively. We also show the shot noise and one-halo term for all the frequencies in Fig.\,\ref{fig:cibfit}. 

In Fig\,\ref{fig:lens_fit}, we show the comparison between our best fit model and the CIB-CMB lensing cross-correlation points. We consider these data points to calculate the best fit of our model. We find a good agreement between the data points and the best fit.

\begin{table}[ht]
 \centering
\begin{tabular}{M{2.5cm}M{2.5cm}}
\hline
\hline 
Parameter &  Marginalised value \\ 
\hline
 $\alpha$      & $0.007\substack{+0.001 \\ -0.001}$ \\[6pt]
 $\beta$       & $3.590\substack{+0.324 \\ -0.305}$ \\[6pt]
 $\gamma$      & $2.453\substack{+0.128 \\ -0.119}$ \\ [6pt]
 $\delta$      & $6.578\substack{+0.501 \\ -0.508}$ \\ [6pt]
 $b_0$        & $0.830\substack{+0.108 \\ -0.108}$ \\ [6pt]
 $b_1$        & $0.742\substack{+0.425 \\ -0.471}$ \\ [6pt]
 $b_2$        & $0.318\substack{+0.275 \\ -0.236}$ \\ [6pt]
 $f_{217}^{cal}$    & $1.000\substack{+0.002 \\ -0.002}$ \\ [6pt]
 $f_{353}^{cal}$    & $1.004\substack{+0.007 \\ -0.007}$ \\ [6pt]
 $f_{545}^{cal}$    & $0.978\substack{+0.015 \\ -0.014}$ \\ [6pt]
 $f_{857}^{cal}$    & $1.022\substack{+0.023 \\ -0.023}$ \\ [6pt]
 $f_{3000}^{cal}$   & $1.155\substack{+0.086 \\ -0.089}$ \\ [6pt]
 \hline
\end{tabular}
\newline
\centering \caption{Marginalised values of the CIB model parameters provided at a 68 \% confidence level}\label{tab:3}
\end{table}

\section{CIB redshift distribution and star formation history \label{sec:CIB_z_SFRD}}
\subsection{Redshift distribution of CIB anisotropies and mean level}
\begin{figure}
\centering
\includegraphics[width=9cm, height=9cm]{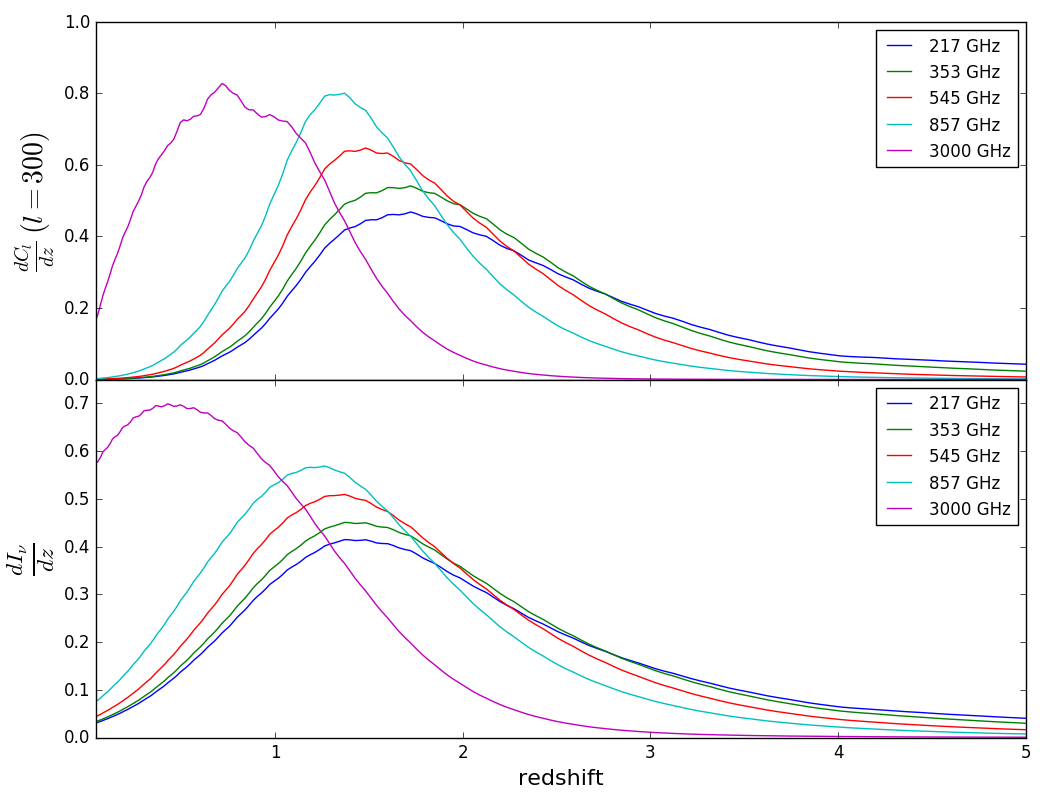}
\centering \caption{Expected CIB mean level and anisotropy redshift distributions are shown in the lower and upper panel, respectively. Both of the distributions have been normalised for easier comparison, and the different colours show different frequency bands. The CIB anisotropy distribution is shown for $\ell=300$.}
\label{fig:red_dist}
\end{figure}

The model we used is based on the SEDs over a range of wavelengths for galaxies which vary with redshift. It is thus interesting to study the redshift distribution of the mean level of the CIB as well as the CIB anisotropies which are based on these SEDs. In Fig.\,\ref{fig:red_dist}, we show the redshift distribution of the CIB anisotropies at $\ell = 300$ and CIB mean level intensity for different frequency bands. Both plots have been normalised, i.e. $\int d(\nu I_\nu)/dz\: dz = 1$ and $\int dC_l/dz\: dz = 1$, so that it is easier to compare the results for different frequencies. \\

\begin{figure}[h!]
\centering
\includegraphics[width=9cm]{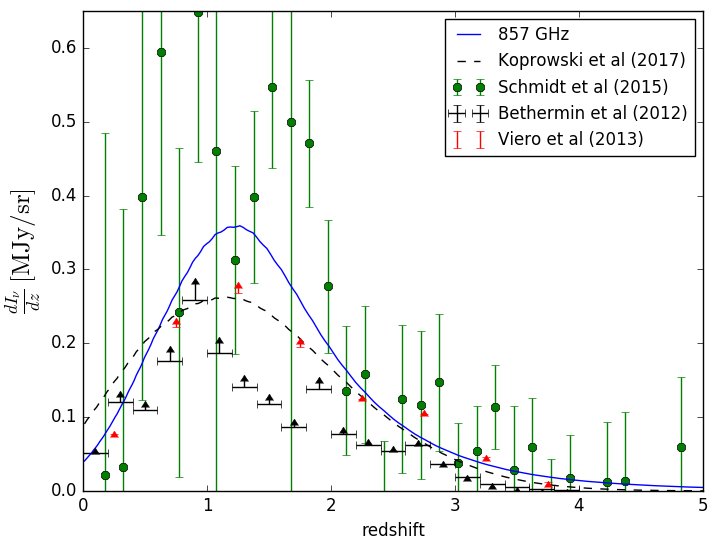}
\centering \caption{Expected CIB mean level redshift distribution at 857\,GHz compared to observational constraints. The lower limits are from \cite{Bethermin_2012_c} and \cite{Viero_2013}. They are shown using black and red triangles, respectively. Measurements from \cite{Schmidt_2015} are shown in green. The redshift distribution derived from \cite{Koprowsky_2017} SFRD constraints is shown with the black dashed line.}
\label{fig:inuobs1}
\end{figure}

It is observed that as we go to lower frequencies, from 3000\,GHz to 217\,GHz, the peak of the CIB anisotropy redshift distribution increases in redshift. This is expected as higher frequencies (lower wavelengths) probe lower redshifts and vice versa \citep[e.g.][]{Lagache_2005, Bethermin_2013}. The CIB mean level distribution follows the same trend. A similar pattern is observed for the CIB anisotropy and for the mean level distribution in \cite{Bethermin_2013} model. We note, however, that the peak of CIB anisotropies predicted in \cite{Bethermin_2013} at 353 and 217\,GHz is reached around z$=$2.4 and the same is reached at lower redshift (around z$=$1.7) for our model.\\
In Fig.\,\ref{fig:inuobs1}, we compare our redshift distribution on the CIB mean level with the lower limits  from \cite{Bethermin_2012_c} and \cite{Viero_2013} which were derived by stacking the 24$\mu$m-selected and mass-selected sources, respectively.  It is observed that the CIB mean level from our model is  higher than these two lower limits for most of the values. We also compare our results with redshift distribution of the CIB derived by \citet{Schmidt_2015} using the cross-correlation between the {\it Planck} HFI maps and SDSS DR7 quasars. It can be seen that although some of the data points from \citet{Schmidt_2015} are consistent with our curve, the measurements tend to be higher than our model at z$>$0.5. A similar trend is followed at other frequencies, and so they are not shown here. \cite{Koprowsky_2017} assumed a Gaussian distribution for the IR and UV $\rho_\textrm{SFR}$ and also provided the best fit values for the distribution. Based on this $\rho_\textrm{SFR}$ form, we calculated the corresponding mean CIB level distribution using Eq. \ref{eq:jnu} and Eq. \ref{eq:clcib} using the SED templates from \cite{Bethermin_2017} (see  Fig.\,\ref{fig:inuobs1}, dashed black curve). It can be  seen that their mean CIB level distribution is lower than some lower limits (at z$\sim$1 and z$\sim$3). It is also lower than values from our model between 0.5$<$z$<$2, and hence in even more tension with the values from \citet{Schmidt_2015}. 

\subsection{Star formation history \label{sec:SFRD}}

\begin{figure}[h]
\centering
\includegraphics[width=9cm]{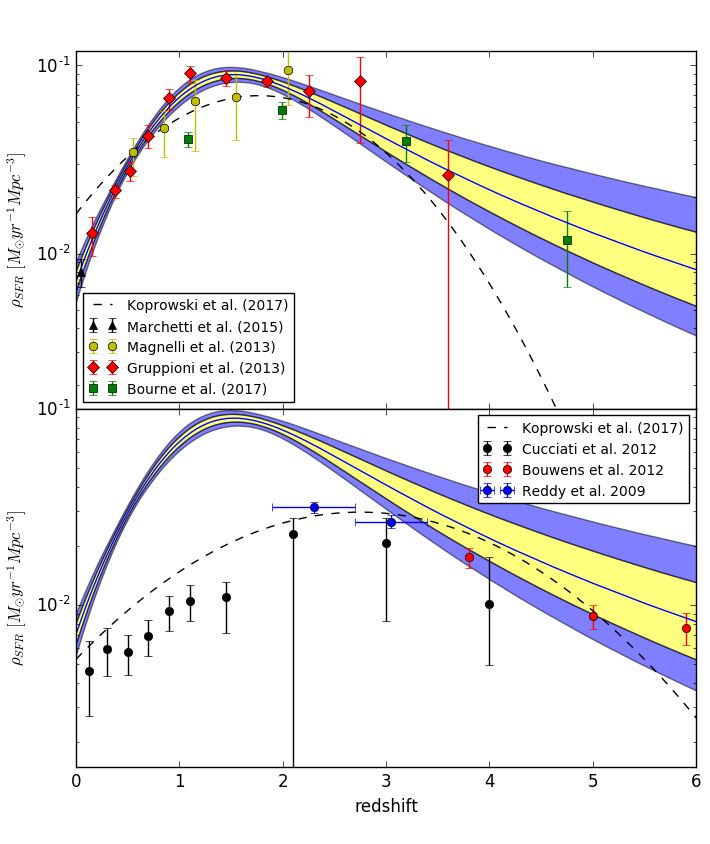}
\centering \caption{Evolution of star formation density with redshift as constrained by the linear CIB model. The $\pm 1\sigma$ and $\pm 2\sigma$ confidence region around the median realisation is shown in yellow and blue,  respectively. Measurements of obscured star formation density from \cite{Gruppioni_2013}, \cite{Magnelli_2013}, and \cite{Marchetti_2015}, which were used to fit the CIB model, have been added along with the measurements from \cite{Bourne_2017} in the upper panel. The Gaussian form of the $\rho_\textrm{SFR}$ for the IR obscured star formation rate density used by \cite{Koprowsky_2017} has been plotted as the black dashed line in the upper panel with a corresponding UV part in the lower panel. Unobscured star formation rate density derived from UV from  \cite{Cucciati_2012}, \cite{Bouwens_2012}, and \cite{Reddy_2009} are also shown in the lower panel.}
\label{fig:rhosfr}
\end{figure}

We show the evolution of star formation density with the redshift in Fig.\,\ref{fig:rhosfr} with the best fit as well as the corresponding $1\sigma$ and $2\sigma$ confidence regions. We created these confidence regions using the chains from the Monte Carlo sampling of our likelihood. We took random samples from the chains for the star formation density parameters and constructed an array of $\rho_\textrm{SFR}$ with these samples for all the redshifts. The median of these values at every redshift is then the central value, and we took samples within 68.2\% and 95.4\% around the central value as $1\sigma$ and $2\sigma$ regions, respectively.\\
We also show the IR measurements from \cite{Gruppioni_2013}, \cite{Magnelli_2013}, \cite{Marchetti_2015}, and \cite{Bourne_2017}, and UV measurements from \cite{Cucciati_2012}, \cite{Bouwens_2012}, and \cite{Reddy_2009},  in the upper and lower panels, respectively.  Our best fit passes through the IR data points considered in the fit. As mentioned in Sect.\,\ref{subsub:extobs}, these data points were obtained using a Salpeter IMF, whereas in our study, we use a Chabrier IMF. Therefore, both the IR and UV data points have been converted to take into account this change.\\
As expected, the UV data points (which have not been corrected for dust attenuation) for unobscured star formation density lie below our curves for $z < 3$ as most of the UV light emitted by young and short-lived stars in this regime is reprocessed by dust. Although the IR contribution is still dominant below $z<4$, the contribution from the UV becomes significant and roughly equal to IR for $z>4$. Hence IR contribution alone is not a good measure of the total star formation rate density at such high redshifts.\\
We plot the best fit IR and UV $\rho_\textrm{SFR}$ curves from \cite{Koprowsky_2017} with dashed black lines in the upper and lower panels of Fig.\,\ref{fig:rhosfr}, respectively. It can be seen that their value of the IR SFRD in the local Universe is higher compared to the other measurements. Also, their IR SFRD drops very quickly at higher redshifts and is basically negligible for z$>$4. Their results are clearly discrepant with ours in these two regimes. A similar trend is followed by the UV SFRD which drops quickly and is much smaller than current observational constraints at z$\sim$6.\\ 
To investigate the discrepancy between our values  and the \cite{Koprowsky_2017} $\rho_\textrm{SFR}$, we performed a MCMC fitting for the CIB anisotropy model (as described in Sect.\,\ref{subsub:fitting}) and fitting for the effective bias and the calibration factors, but fixing the SFRD to the best fit of \cite{Koprowsky_2017}. We found the effective bias at redshifts z$>$2.5 to be much higher ($>$50\%) than is observed for the SMGs. This effective bias value is too high compared to what is realistically expected, suggesting that their SFRD is underestimated.\\
Similarly, fig.\,\ref{fig:rhosfr} shows that the \cite{Bourne_2017} points at z$\simeq$1 and 2 are a factor $>$2 and 1.3 below our measurements, respectively. Taking these points as priors for $\rho_{SFR}$ rather than those from \cite{Gruppioni_2013} would give a lower measurement for SFRD and this would affect the measurement of the linear bias. We investigated the consistency of the solution obtained on both $\rho_\mathrm{SFR}$ and $b_\mathrm{eff}$ either by taking the \cite{Bourne_2017} data points as priors or by forcing and fixing the SFRD to go through the \cite{Bourne_2017} data points in our MCMC analysis. We found that the linear bias is severely overestimated compared to similar galaxy populations, for example by a factor $>3$ at z$\simeq$1.2 compared to the SMG sample of \cite{Wilkinson_2017}. This shows that the SFRD determined by \cite{Bourne_2017} at low $z$ seems to be underestimated.

\section{Host dark matter halos of the CIB \label{sec:DMH_Mass}}

\subsection{Effective bias of CIB galaxies}

\begin{figure}[h]
\centering
\includegraphics[width=9cm]{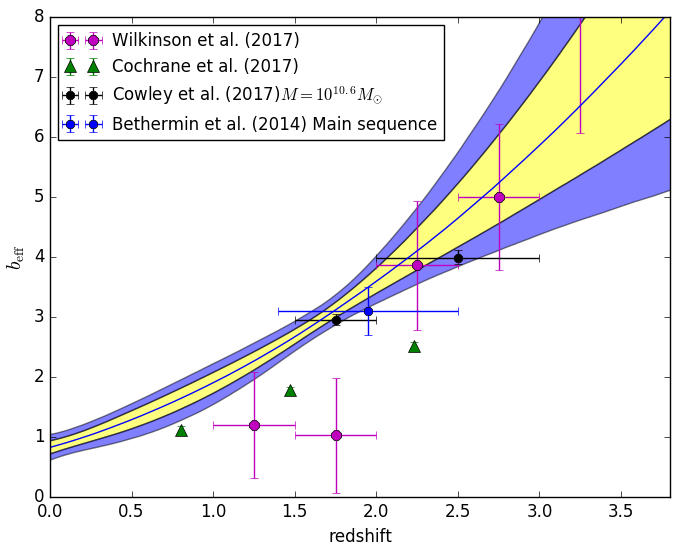}
\centering \caption{Evolution of the effective bias with redshift derived from the CIB compared to the observational values obtained on selected populations of galaxies: \cite{Bethermin_2014}, \cite{Cochrane_2017}, \cite{Cowley_2017}, and \cite{Wilkinson_2017} using a sample of BzK-selected galaxies, $\textrm{H}_\alpha$-selected galaxies, mass-selected galaxies, and SMGs, respectively.}
\label{fig:biasnorm}
\end{figure}

We also study the evolution of the effective bias and find that it is increasing with redshift, as expected. In Fig.\,\ref{fig:biasnorm}, we compare our measurements with the clustering measurements of individual galaxies selected using various criteria. 

At z$>$2, our measurements and the clustering of submillimetre galaxies (SMGs) by \citet{Wilkinson_2017} are very similar. SMGs are the most star-forming galaxies at z$\gtrsim$2 \citep[e.g.][]{Blain_2002,Chapman_2005}, and thus it is not completely surprising to find a similar bias since a large fraction of the SFRD at z$>$2 is hosted by galaxies forming more than 100\,M$_\odot$/yr \citep[e.g.][]{Caputi_2007,Magnelli_2009,Bethermin_2011,Gruppioni_2013}. At z$<$2, \citet{Wilkinson_2017} found a bias close to 1, which is 1\,$\sigma$ and 2\,$\sigma$ below our model in the 1$<$z$<$1.5 and 1.5$<$z$<$2 bins, respectively. This might be a statistical fluctuation coming from the small sample in their low-z bins ($\sim$60 objects). In addition, the algorithm they use to associate SMGs with optical counterparts and photometric redshift is only reliable at 83\,\%, which could induce biases in the clustering measurements. In the 1.5$<$z$<$2 bin, if they consider only objects with a solid radio identification of the counterparts, they find a bias of 1.65$\pm$1.09, which is compatible at 1\,$\sigma$ with our measurements, which means that  there might be no real tension in these low-z bins.

We also compare our bias estimates with selections at other wavelengths. \citet{Bethermin_2014} measured the clustering of massive star-forming galaxies at z$\sim$2 selected using the BzK color criterium \citep{Daddi_2004}. Our effective bias agrees with this measurement. \cite{Viero_2013} show that majority of the contribution to the CIB is emitted by massive dusty star-forming galaxies with log$(M/M_{\odot})$ 10.0--11.0.  We plot the clustering of the mass-selected (M$_\star = 10^{10.6}$\,M$_\odot$) galaxies measured by \cite{Cowley_2017} and find an agreement with our measurements, as expected. \citet{Ishikawa_2015} measured the clustering of gzKs-selected galaxies for various depths and found a strong dependance of bias with the depth. Their deep sample (K$\le$23) has a much smaller bias than the CIB (1.8 versus 3.1), but the shallowest sample (K$\le$21) has a stronger bias (4.16). The CIB measurements are indeed dominated by the galaxies, which contribute the most to the star formation budget, while clustering measurements of galaxy populations are dominated by the most numerous objects, which are usually the numerous low-mass objects residing in low-mass halos (M$_{\rm halo} < 10^{12}$\,M$_\odot$) with a lower clustering. It is thus expected that the deepest K-band sample with masses and SFRs well below the knees of the mass and SFR functions have a lower bias than the CIB. In a similar way, the bias of H$_\alpha$-selected galaxies, which are mainly low-SFR and low-mass galaxies, measured by \citet{Cochrane_2017} (1.12, 1.78, 2.52), is significantly lower than our model (1.64,  2.62,  4.06) at  redshifts 0.8,  1.47,   2.23 respectively.

\subsection{What does the effective bias mean? \label{sect:beff}}
Throughout the paper, we have used the effective bias term ($b_\textrm{eff}$) and it should be noted that effective bias might not be equal to the mean bias of halos hosting the dusty galaxies ($b(M,z)$). As shown in Appendix C of \cite{Planck_cib_2014}, the effective bias is the mean bias of halos hosting the dusty galaxies weighted by their differential contribution to the emissivities and is given by the following equation:
\begin{equation} \label{eq:beff_expl}
b_{eff} = \dfrac{\int \dfrac{dj}{dM}(\nu,z)b(M,z)dM}{\int \dfrac{dj}{dM}dM}\,.
\end{equation} 
In this equation, $\dfrac{dj}{dM}$ is the differential contribution of a range of halo mass to the emissivity. Since the emissivity is directly proportional to $\rho_{SFR}$, we can rewrite this equation as
\begin{equation} \label{eq:beff_expl2}
b_{eff} = \dfrac{\sum\limits_{vol}\mathrm{SFR}\times b(M,z)}{\sum\limits_{vol}\mathrm{SFR}}\, ,
\end{equation} 
where we sum over all the halos and their host galaxies in a given volume. Here $b(M,z),$ which is the bias as a function of halo mass and redshift, is calculated with the formula which has been calibrated using the numerical simulations from \cite{Tinker_2010},
\begin{equation} \label{eq:b_m_z}
b(\nu) = 1 - A\dfrac{\nu^a}{\nu^a + \delta_c^a} + B\nu^b + C\nu^c \,,
\end{equation}
where $\nu = \delta_c/\sigma_M$  characterises the peak heights of the density field as function of mass and redshift; $\delta_c$ is the linear critical density at a given redshift for a given cosmology; and $\sigma_M$ is the linear matter variance in a top hat filter of width $R = (3M/4\pi\bar{\rho_0})^{1/3}$ with M and $\bar{\rho_0}$ being the mass and mean density for a given halo, respectively. Details of these calculations have been given in  Appendix A of \cite{Coupon_2012}.  $A,a,B,b,C,\: \textrm{and}\: c$ are functions of $\Delta$ which is the overdensity for a given halo defined as the mean interior density for the halo relative to the background. These functions are provided in Table\,2 in \cite{Tinker_2010}. As often used, we took a virial value of $\Delta \approx 200$. From Eq. \ref{eq:beff_expl}, we can see that if halos in a given mass range contribute more or less to the emissivity compared to halos of a different mass range, there will be a slight offset between b$_\textrm{eff}$ and b(M,z). \\
It is interesting to quantify this offset, particularly in the context of the measurement of the mass of the typical host dark matter halos which contribute to the CIB, and hence to the obscured star formation. This is the purpose of the next section.\\

As a first approximation, we can estimate the mean host halo mass at a given redshift by equating $b_\textrm{eff}$ and $b(M,z)$. At $z = 2$, we find that the mass of the dark matter halos which contribute the most to the CIB calculated this way is $\log_{10}M = 12.93\substack{+0.084 \\ -0.081}$, where 0.084 and 0.081 are the 1$\sigma$ upper and lower limits, respectively.

\subsection{Host dark matter halo mass of the dusty star formation} \label{sect:mhalo_corr}
As is shown in the previous section, $b_\textrm{eff}$ might not be exactly  equal to the mean bias of halos hosting the dusty galaxies ($b(M,z)$). 
To understand how this effective bias relates to the bias of the halos contributing the most to the star formation budget, we used the Simulated Infrared Dusty Extragalactic Sky (SIDES) simulation \citep{Bethermin_2017}, where the impact of clustering and angular resolution for far-infrared and millimetre continuum observations has been taken into account. \\

\begin{figure}[h]
\centering
\includegraphics[width=9cm]{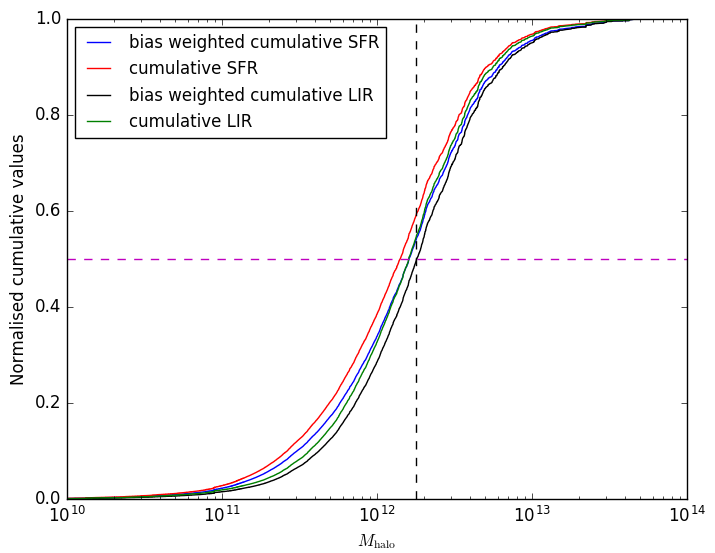}
\centering \caption{Normalised cumulative SFR density and LIR density as a function of host dark matter halo mass for galaxies in SIDES simulation are shown in red and green, respectively. We also show the variation of these quantities when they are weighted with the corresponding bias value for the galaxies in blue and black, respectively. The black vertical line represents the mass of the dark matter halo mass calculated converting the $b_\mathrm{eff}$ obtained using the LIR instead of the SFR into the halo mass using Eq.\,\ref{eq:b_m_z} and Eq.\,\ref{eq:beff_expl2} (replacing SFR by LIR). The horizontal line represents the 50\% cumulative level. This plot has been made for the galaxies  in the range  $2.0<z<2.2$.}
\label{fig:cumsfr}
\end{figure}

The SIDES simulation provides the star formation rate for galaxies corresponding to a given halo mass M$_{halo}$ at a given redshift z. In Fig.\,\ref{fig:cumsfr} we plot the cumulative star formation rate as a function of halo masses for galaxies with $2.0<z<2.2$. We can see from the red curve in Fig.\,\ref{fig:cumsfr} how the cumulative SFR density grows with the halo masses in SIDES. We define the mass at which the above curve reaches 50\% of the total cumulative star formation as the dark matter halo mass hosting the dusty galaxies. 

We calculated the bias b(M,z)  corresponding to each halo at a given redshift in SIDES using Eq.\,\ref{eq:b_m_z}. The blue curve in Fig.\,\ref{fig:cumsfr}, shows how the normalised cumulative SFR$\times$b(M,z) grows with the halo mass. It is seen that 50\% of the total SFR or SFR$\times$b is reached around a halo mass of 10$^{12}$\,M$_\odot$, but not at the exact same halo mass. Since the bias is higher for massive halos hosting more massive and star-forming galaxies, we reach these 50\% for SFR$\times$b at a mass 0.06\,dex higher than for just SFR. The effective bias of the CIB, thus corresponds to a bias slightly higher than the bias of the pivot halo mass at which we have 50\,\% of the SFRD. This value of 0.06\,dex, however, is still not the offset we should apply to the mass $\log_{10}M = 12.93\substack{+0.084 \\ -0.081}$ of the host dark matter halo obtained using $b_\mathrm{eff}$.\\

The CIB  traces only the obscured star formation, but at low stellar mass (and thus halo mass), a significant fraction of the UV from the star formation escapes the galaxies. We used the method of \citet{Bernhard_2014} based on the empirical relation between stellar mass and dust attenuation of \citet{Heinis_2014} to predict the SFR$_{\rm IR}$, i.e. the obscured star formation, from which we deduce an L$_{\rm IR}$ (obscured IR luminosity) for all the sources in the simulation. In Fig.\,\ref{fig:cumsfr}, the green and black curves represent the cumulative contribution of the L$_{\rm IR}$ and the  L$_{\rm IR}\times$b, respectively. This latest quantity (L$_{\rm IR}\times$b) is the closest to what we actually measure with the CIB. The halo mass at which 50\,\% of the total quantity is reached for slightly higher halo mass (0.05 \,dex) than that found for SFR$\times$b because obscured star formation is slightly biased toward massive galaxies (and thus halos). For the dark matter halos between $2 < z < 2.2$, we observe a difference of around 0.11\,dex between halo masses found using only the SFR and the bias weighted LIR (L$_{\rm IR}\times$b). Using Eq. \ref{eq:beff_expl2}, we can calculate the $b_\textrm{eff}$ corresponding to these LIR values where we just replace the SFR term in the equation with the LIR obtained here. Using the procedure followed in Sect.\,\ref{sect:beff}, we convert this $b_\textrm{eff}$ to the corresponding halo mass using Eq.\,\ref{eq:b_m_z}. The black vertical line in Fig.\,\ref{fig:cumsfr} represents this mass of the dark matter halo. Finally, the difference between this mass and halo mass found using only the SFR, which is 0.10 dex, is the actual offset we are looking for at this redshift bin. \\ 

\begin{figure}[h]
\centering
\includegraphics[width=9cm]{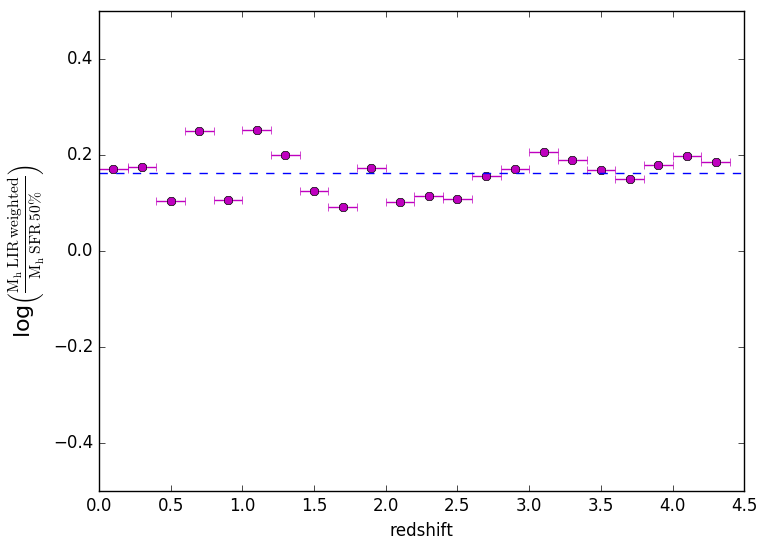}
\centering \caption{Difference between the dark halo mass estimated as the halo mass where 50\% cumulative SFR (M$_h$ SFR 50\%) is achieved and the halo mass calculated by converting the b$_{eff}$, which is obtained using the LIR (replacing SFR with LIR in Eq. \ref{eq:beff_expl2}) to the corresponding halo mass using Eq. \ref{eq:b_m_z} (M$_h$ LIR weighted). This difference has been calculated and is shown here for all the redshift bins. The dashed line shows the mean of these values: around 0.16 dex. This is the offset  applied to the halo mass derived from b$_{eff}$ to obtain the mean mass of dark matter halos contributing to CIB.}
\label{fig:mhalodiff}
\end{figure}

Following the procedure mentioned above, we calculate this mass offset for all the redshift bins. This offset is shown in Fig.\,\ref{fig:mhalodiff}. The dashed blue line in the figure shows the mean value of the offset over all the bins which is around 0.16 dex. It is clear from the figure that this offset is not constant at all redshifts. After applying this mean correction to the halo masses at all the redshifts, we also propagate the uncertainty on this offset to the error bars on the mass measurements. We find that the mass of the dark matter halos which contribute the most to the CIB at $z=2$ comes out to be $\log_{10}M = 12.77\substack{+0.128 \\ -0.125}$, where 0.128 and 0.125 are the 1$\sigma$ upper and lower limits, respectively (considering additional uncertainty on mass offset), compared to the previous value of $\log_{10}M = 12.93\substack{+0.084 \\ -0.081}$. \\

\begin{figure}[h]
\centering
\includegraphics[width=9cm]{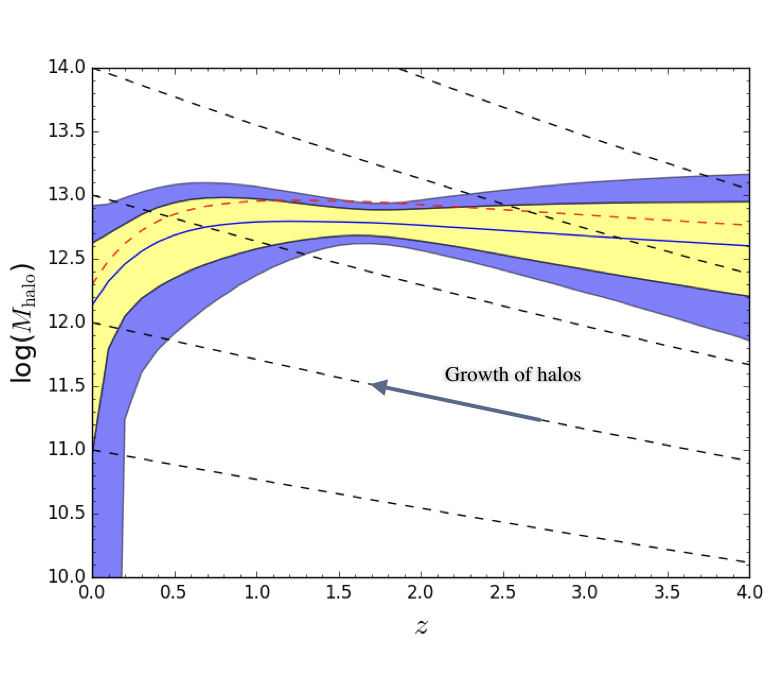}
\centering \caption{Mass of the dark matter halos hosting the galaxies contributing to the CIB as a function of redshift. The black dashed lines show the growth of the dark matter halo mass with redshift. The red dashed line shows the mass that would be obtained directly equating $b_\textrm{eff}$ and $b(M,z)$ (see Sect.\,\ref{sect:beff}).}
\label{fig:b_m_z_corr}
\end{figure}

We show in Fig.\,\ref{fig:b_m_z_corr} the variation of the host halo mass as a function of redshift for $z\le4$. Contours represent the $1\sigma$ and $2\sigma$ confidence regions. We obtain an almost constant dark matter halo mass ($\approx 10^{12.7}$M$_\odot$) contributing to the CIB from 1$<$z$<$4,  and then it starts decreasing from z$<$1.
Host halos grow over time and the dashed lines show this growth of halos with redshift, as computed by \cite{Fakhouri_2010}. We see that most of the star formation at z $>$ 2.5 occurred in the progenitors of clusters (M$_h(z=0) > 10^{13.5}$M$_\odot$). Then at the lower redshifts from 0.3$<$z$<$2.5, most of the stars were formed in groups ($10^{12.5}$ < M$_h(z=0) < 10^{13.5}$M$_\odot$) and later on, finally, inside the Milky Way-like halos ($10^{12}$ < M$_h(z=0) < 10^{12.5}$M$_\odot$) for z$<$0.3. Although a bit high, these results are compatible with the results obtained by \citet{Bethermin_2013}.

\section{Effect of cosmology on CIB \label{sec:cosmo_cib}}
\subsection{Method}
All the previous studies on the CIB were performed assuming a fixed fiducial cosmology. We can see the role of the cosmological parameters for the  CIB through Eq.\,\ref{eq:clcib} where the CIB power spectrum depends upon the dark matter power spectrum, Hubble's constant, cosmological energy density parameters (through distance measures), etc. It was thus  interesting to study the effect of cosmology on the CIB, i.e. whether changing the cosmology makes a significant impact on the CIB parameters. In order to study this effect, we performed a Fisher matrix analysis over all  12 CIB parameters and 6 cosmological parameters. \\

For an N-variate multivariate normal distribution $X \sim N(\mu(\theta), \Sigma(\theta))$, the $(m, n)$ entry of the Fisher matrix is given as
\begin{equation} \label{eq:Fisher}
I_{m, n} = \dfrac{\partial \mu}{\partial \theta_m} \Sigma^{-1} \dfrac{\partial \mu^T}{\partial \theta_n} + \dfrac{1}{2} \textrm{tr} \Big(\Sigma^{-1} \dfrac{\partial \Sigma}{\partial \theta_m} \Sigma^{-1} \dfrac{\partial \Sigma}{\partial \theta_m} \Big),
\end{equation}
where $()^T$ denotes the transpose of a matrix and $\textrm{tr}()$ denotes the trace of a square matrix; $\theta = \big[ \theta_1, ... , \theta_K \big]$ is a $K$-dimensional vector of parameters being considered; $\mu(\theta) = \big[\mu_1(\theta), ... , \mu_N(\theta) \big]$ are the mean values of the N random variables for which the uncertainties on their measurements are available;  $\Sigma(\theta)$ is the covariance matrix for all the variables $\mu(\theta)$; and 
\begin{equation}
\dfrac{\partial \mu}{\partial \theta_m} = \Big[ \dfrac{\partial \mu_1}{\partial \theta_m}, ... , \dfrac{\partial \mu_N}{\partial \theta_m}\Big]
\end{equation}
is a vector of the partial derivatives of the N variables for a given parameter $\theta_m$. In our case, $\Sigma(\theta) = \textrm{constant,}$ i.e. the covariance matrix is independent of the parameters and hence the second term in the Eq.\,\ref{eq:Fisher} vanishes as $\dfrac{\partial \Sigma}{\partial \theta_m} = 0,$ and therefore
\begin{equation} \label{eq:fish}
I_{m, n} = \dfrac{\partial \mu}{\partial \theta_m} \Sigma^{-1} \dfrac{\partial \mu^T}{\partial \theta_n} \,.
\end{equation}
In our case, the parameters being considered are\\
 $\theta = \big[ \alpha, \beta, \gamma, \delta, b_0, b_1, b_2, f_{\nu}^\textrm{cal}, H_0,  \Omega_bh^2, \Omega_ch^2, \tau,$ $n_s, A_s \big]$. \\
 In the following sections, we explain the construction of the $\dfrac{\partial \mu}{\partial \theta}$ and $\Sigma$ matrices.\\

\subsection{$\dfrac{\partial\mu}{\partial\theta}$ matrix construction} 
\begin{figure}[h]
\centering
\includegraphics[width=9cm]{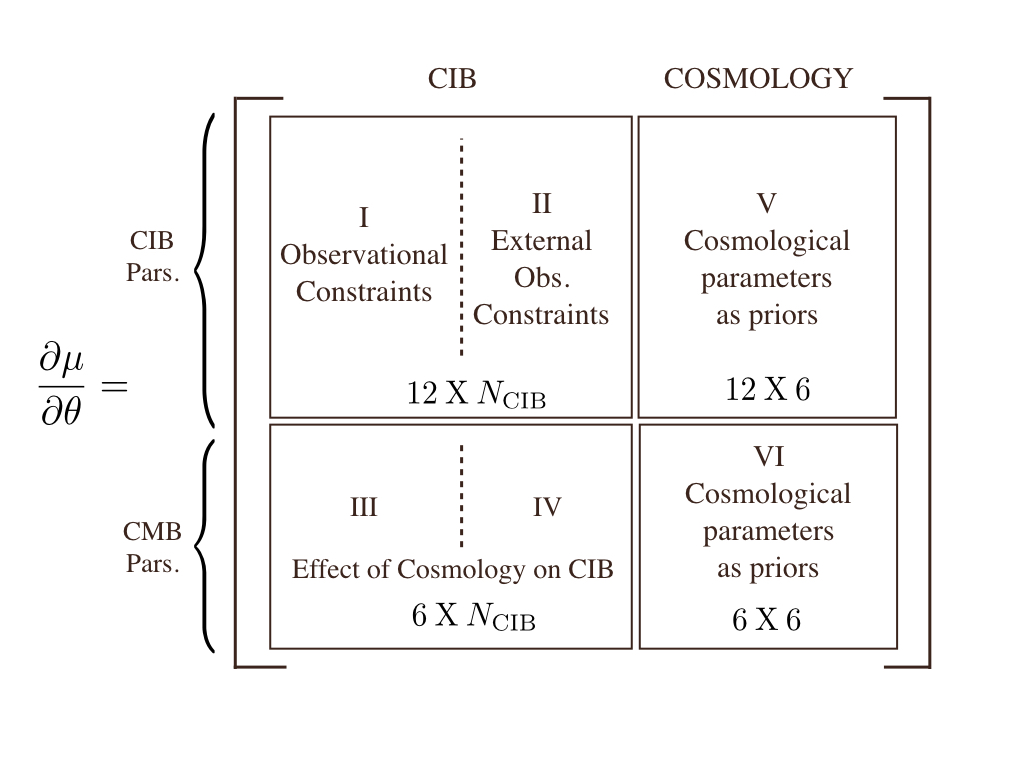}
\caption{Details of the matrix $\dfrac{\partial\mu}{\partial\theta_m}$ used in the Fisher matrix analysis.}
\label{fig:fish1}
\end{figure}

Figure\,\ref{fig:fish1} shows the different components of the $\dfrac{\partial \mu}{\partial \theta}$ matrix. To construct this matrix, we have to calculate the partial derivatives of all the variables with respect to all the CIB and cosmological parameters. For every parameter $\theta_m$, the partial derivative is calculated as
\begin{equation}
\dfrac{\partial\mu}{\partial\theta_m} = \dfrac{\mu(\Theta, \theta_m + \delta\theta_m) - \mu(\Theta, \theta_m - \delta\theta_m)}{2\delta\theta_m}
,\end{equation}
where while calculating the partial derivative for a parameter $\theta_m$, all the other parameters represented here by $\Theta$ are kept constant at their best fit value. As we are performing the Fisher analysis to see the relative effect of cosmological parameters on CIB, we assume a fiducial cosmology and treat all the cosmological parameters as priors. We maintain the value of $\delta\theta$ small enough to ensure that we are within the Gaussian region for a given parameter. \\
Component I in the upper left part of the matrix contains the observational constraints mentioned in Sect.\,\ref{subsub:obs}, i.e. CIB auto- and cross-power spectra, priors on calibration factors $f_\nu^\textrm{cal}$, and CIB-CMB lensing cross-correlation points for all the frequencies. Component II in the same upper left part of the figure contains the external observational constraints mentioned in Sect.\,\ref{subsub:extobs}, which are the $\rho_\textrm{SFR}$ observations, prior on local bias, and mean level of CIB at different frequencies. It should be noted that all the $\rho_\textrm{SFR}$ values have been calculated by different groups using different cosmologies. In order to perform the Fisher analysis, we needed to make them cosmology independent and therefore we converted them to observed flux values using Eq.\,\ref{eq:flux} assuming a single fiducial Planck 2015 cosmology for all the points. Components III and IV in the lower left part of Fig.\,\ref{fig:fish1} calculate the effect of the six cosmological parameters on the CIB variables (observational constraints in component III and external observational constraints in component IV). \\
Calculations in some of the cases are simplified as certain variables are dependent on only one of the parameters, for example the Gaussian prior on the calibration factor for a given frequency $f_{\nu}^\textrm{cal}$ is independent of all the parameters except itself, and its partial derivative with respect to itself is one. Therefore, the column vector corresponding to this variable in the matrix would simply be \{0,0,...,1,0,...,0\}. \\
Component V in the upper right part of the matrix calculates the effect of CIB parameters on six cosmological parameters. As mentioned above, we treat cosmological parameters as priors, hence they are independent of the CIB parameters, and their partial derivatives with respect to CIB parameters is zero. Therefore, this part of a matrix is an array of the size 12 x 6 with all the elements being zero. Similarly, component VI of the matrix is a diagonal matrix with all the diagonal entries being one and all the off-diagonal elements being zero. 

\begin{figure}[h]
\centering
\includegraphics[width=9cm]{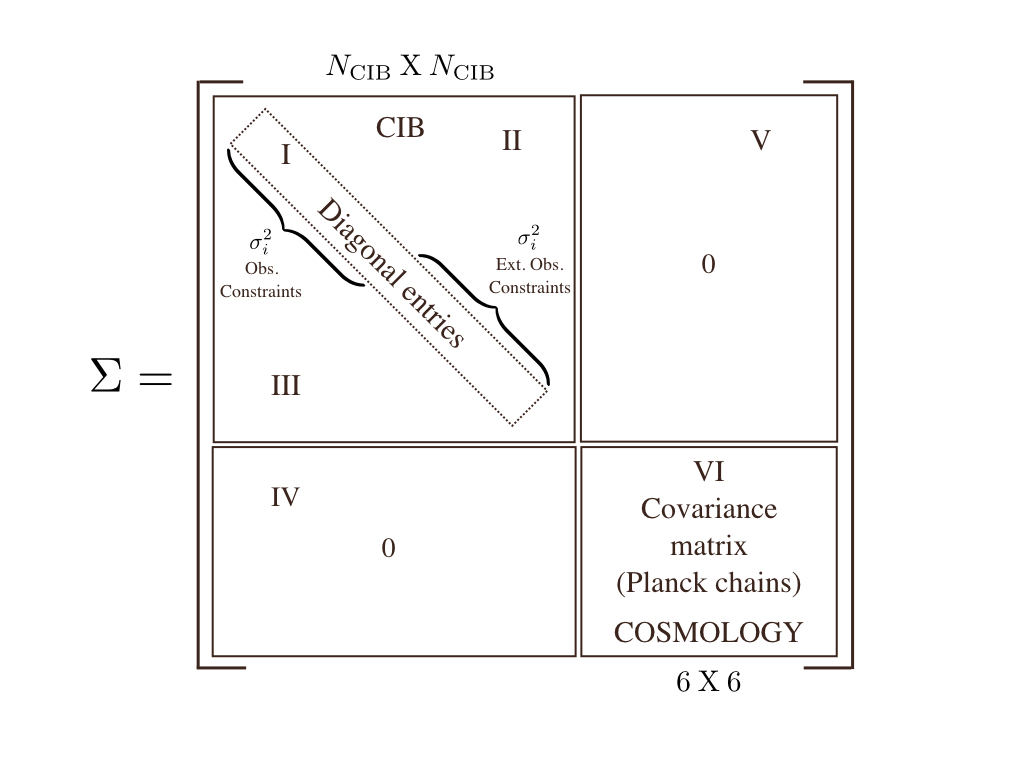}
\caption{Details of the matrix $\Sigma$ used in the Fisher matrix analysis.}
\label{fig:fish2}
\end{figure}

\subsection{$\Sigma$ matrix construction}
The $\Sigma$ matrix is the covariance matrix and it contains the information of the error bars for the variables used in the $\dfrac{\partial\mu}{\partial\theta}$ matrix. As mentioned in Sect.\,\ref{subsub:fitting}, we assume Gaussian uncorrelated error bars for the data points and this makes it relatively easy to construct this matrix. Its elements are shown in Fig.\,\ref{fig:fish2}. The upper left panel of the figure contains the covariance matrix from the CIB variables, which contains  the observational and the external observational constraints. As there is no correlation between different CIB variables, the upper left panel is just a diagonal matrix with component I containing the square of the error bars for every variable. The off-diagonal components II and III are both zero. We  treat the cosmological parameters as priors. Cosmological parameters are very well constrained by {\it Planck} \citep[see][]{Planck_cosmo_2016} and the Monte Carlo sampling chains for their likelihood is provided on the Planck Legacy Archive \footnote{\url{https://pla.esac.esa.int/pla/}}. We used these chains to calculate the covariance matrix between the six cosmological parameters. This matrix goes in component VI of Fig.\,\ref{fig:fish2}. Components IV and V are matrices with all the elements being zero. \\
Once we have $\dfrac{\partial\mu}{\partial\theta}$ and $\Sigma$ matrices, it is straightforward to calculate the entries of the Fisher matrix with Eq.\,\ref{eq:fish}; in our case,  has the dimensions of (18,18).   

\subsection{Results}
We show the results from our Fisher matrix analysis in Table\,\ref{tab:fisher}. The error bars on the CIB parameters change very little when we add the effect of varying the cosmological parameters. The change in the error bars on the parameters varies from 0.04\% to 3.4\% which is within the uncertainties on the CIB parameters. Therefore, we conclude that the cosmological parameters determined at the current level of precision have a negligible impact on the CIB model.  

\begin{table}[h]
 \centering
\begin{tabular}{cccc}
\hline
\hline 
Parameter &  \begin{tabular}{@{}c@{}}$1\sigma$ Error \\ (w/o cosmology\\ information)\end{tabular}   &  \begin{tabular}{@{}c@{}}$1\sigma$ Error \\ (with cosmology\\ information)\end{tabular} & \% change \\ 
\hline
 $\alpha$      & 0.00087 & 0.00087 & 0.0715 \\
 $\beta$       & 0.3217 & 0.3218 & 0.0447 \\ 
 $\gamma$      & 0.1415 & 0.1416 & 0.0806 \\ 
 $\delta$      & 0.5847 & 0.5892 & 0.7833 \\
 $b_0$        & 0.8018 & 0.8291 & 3.414 \\ 
 $b_1$        & 1.246 & 1.268 & 1.772 \\ 
 $b_2$        & 0.4345 & 0.4397 & 1.189 \\ 
 \hline
\end{tabular}
\newline\newline
\centering \caption{Fisher analysis results for the CIB parameters. This table gives the error on the CIB model parameters and $f_{\nu}^{cal}$ taking into account or not the errors on the cosmological parameters. }\label{tab:fish_res}
\label{tab:fisher}
\end{table}

\section{Conclusion \label{sec:cl}}
We have developed a linear CIB model and used conjointly the CIB anisotropies and CIBxCMB lensing cross-correlations measured at large scale by {\it Planck} to determine the SFRD up to z=6 and the evolution of the effective bias. Our paper improved upon the analysis of the linear CIB model performed by \cite{Planck_cib_2014}. We used a functional form for the SFRD \citep{Madau_2014} and a polynomial functional form for the effective bias evolution with redshift. We use effective SEDs derived from the latest observations and modelling \citep{Bethermin_2015, Bethermin_2017}. The inclusion of the CIB-CMB lensing cross-correlation in our fitting helps us to partially break the degeneracy between the effective bias and the SFRD parameters. In order to get better constraints on the SFRD parameters, we also used external observational constraints on the SFRD at different redshifts from different surveys which are converted to observed flux values to account for the different cosmologies used in the analysis. We also used the mean level of the CIB measured at different frequencies to get a better constraint on the CIB model parameters. Gaussian priors have been put on the local value of the effective bias along with the calibration factors for different frequencies where the improved values of the uncertainties on the calibration factors from Planck PR2 data release compared to the PR1 data have been taken into account.\\

With these improved constraints on the CIB model parameters, we derived the SFRD of dusty galaxies up to z=6. We showed that UV SFRD measurements are consistently lower than our IR SFRD measurement for z$<$4 and become compatible (at $1\sigma$) with the IR SFRD at z=5. The effective bias  increases steeply with redshift and is compatible with a number of other measurements, and in particular with the bias obtained from the mass-selected M$_\star = 10^{10.6}$\,M$_\odot$ sample of \cite{Cowley_2017}, as expected from the mass distribution of the CIB \citep{Viero_2013}. The possible deviation in the redshift bin 1.5$<$z$<$2 with the SMG sample of \cite{Wilkinson_2017} might be a statistical fluctuation arising from the small sample size in this redshift bin. Consistency checks performed on the $\rho_\mathrm{SFR}$ and effective bias revealed that the IR SFRD from \cite{Bourne_2017} at z$\simeq$1 and \cite{Koprowsky_2017} at z$>$3.5 are underestimated. We found that the redshift distribution of the CIB mean was consistently above the published lower limits and that it peaks at $z\sim1.7$ at 353 and 217\,GHz. \\

Having measured the effective bias value, we have estimated the typical host dark matter halo mass of galaxies contributing to the CIB. As the effective bias of the galaxies we measure is weighted by their SFR (and even more specifically by their L$_{IR}$), the value of the host dark matter halo mass obtained using the effective bias value has to be corrected. We have used the SIDES simulations from \cite{Bethermin_2017} to quantify the amplitude of the correction, which is found to be 0.1\,dex. Using this offset, we find that the typical mass of the host halos contributing to the CIB is $\log_{10}M = 12.77\substack{+0.128 \\ -0.125}$ at z=2, which is slightly on the high side of the range $10^{12.1\pm0.5}$ M$_{\odot}$ to $10^{12.6\pm0.1}$ M$_{\odot}$ found by \citet{Viero_2013} and \citet{Planck_cib_2014} but in very good agreement with \cite{Chen_2016} for faint SMGs of $\log_{10}M = 12.7\substack{+0.1 \\ -0.2}$. \\

Finally, we have quantified for the first time the effect of the cosmology on the CIB parameters. All the previous studies on the CIB had been performed using a fiducial background cosmology. We performed a Fisher matrix analysis to study the effect of changing the cosmology on the CIB and found that it is negligible compared to the existing measurement uncertainties on the CIB parameters. 

\begin{acknowledgements}
We acknowledge financial support from the \textquotedblleft Programme National de Cosmologie and Galaxies'' (PNCG) funded by CNRS/INSU-IN2P3-INP, CEA, and CNES, France; from the ANR under the contract ANR-15-CE31-0017; and from the OCEVU Labex (ANR-11-LABX-0060) and the A*MIDEX project (ANR-11-IDEX-0001-02) funded by the \textquotedblleft Investissements d'Avenir'' French government programme managed by the ANR.
Abhishek Maniyar warmly thanks Sylvain De la Torre and Carlo Schimd for the enlightening discussions on the Fisher matrix analysis.
\end{acknowledgements}
\appendix 
\section{Alternative parametric forms for the effective bias}
\begin{figure}[h]
\centering
\includegraphics[width=9cm]{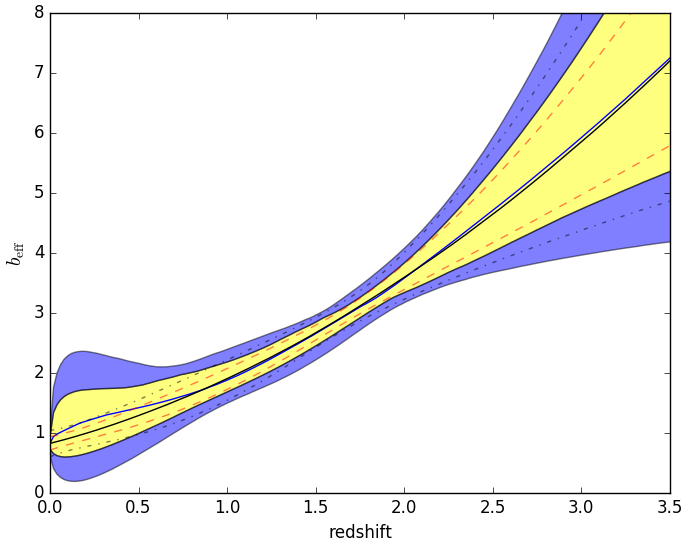}
\caption{Evolution of the effective bias with redshift derived from the CIB using two different parametrisations, as described in  Eq.\,\ref{eq:3bias} (plotted as the blue line with the  $1\sigma$ and $2\sigma$ regions given in yellow and blue, respectively) and Eq.\,\ref{eq:beff} (plotted as the dark line with the $1\sigma$and $2\sigma$ regions given with the red dashed and blue dot-dashed lines, respectively).}
\label{fig:biascomp1}
\end{figure}

As mentioned in Sect.\,\ref{sec:cib_model}, we tested alternative parametric forms for the effective bias with different rates of evolution with redshift compared to $b_{eff}$ used in Eq.\,\ref{eq:beff}. The first alternative model we used calculates $b_{eff}$ following
\begin{equation}\label{eq:3bias}
b_{eff} (z)= b_0 + b_1z^{0.5} + b_2z + b_3z^{1.5} \,.
\end{equation}
The evolution of this effective bias with redshift is shown in Fig.\,\ref{fig:biascomp1}. For comparison, we also show the effective bias from Eq.\,\ref{eq:beff} and its $1\sigma$ and $2\sigma$ regions. We can see that the constraints on $b_{eff}$ become slightly weaker for the new effective bias compared to the original one. This effect is especially strong at lower redshifts where the $1\sigma$ and $2\sigma$ regions on the $b_{eff}$ are much higher with the new parameterisation. \\

Similar to this case, we  also checked the evolution of $b_{eff}$ by adding one more parameter to Eq.\,\ref{eq:3bias} i.e. $b_{eff} (z)= b_0 + b_1z^{0.5} + b_2z + b_3z^{1.5} + b_4z^2$. We get similar results for this case with $b_{eff}$  evolving  faster than the above case and with higher uncertainties. 

\bibliographystyle{aa}
\bibliography{bib2}

\end{document}